\documentclass[aps,prd,superscriptaddress,
onecolumn,showpacs,nofootinbib,preprintnumbers,notitlepage]{revtex4-1}

\usepackage{lipsum}
\usepackage{graphicx}  % needed for figures
\usepackage{multirow}
\usepackage[dvipsnames]{xcolor}
\usepackage[colorlinks=true,allcolors=BlueViolet]{hyperref}
\usepackage[T1]{fontenc}
\usepackage{latexsym}
\usepackage{amsfonts}
\usepackage{amsmath,amssymb}
\usepackage{slashed}
\usepackage{color}
\usepackage{mathtools}
\usepackage{supertabular} 
\usepackage{epsfig}
\usepackage{array}

\usepackage{natbib}
\renewcommand{\arraystretch}{1.2}
\usepackage{xspace}
\usepackage{slashed}
\newcommand{\ie}{{\it i.e.}\xspace}

\DeclareMathOperator{\im}{Im}
\DeclareMathOperator{\re}{Re}

\usepackage{soul,color}
\definecolor{chromeyellow}{rgb}{1.0, 0.65, 0.0}
\definecolor{applegreen}{rgb}{0.55, 0.71, 0.0}

\newcommand{\braket}[3]{\left\langle #1 \right\rvert #2 \left\lvert #3 \right\rangle}

\newcommand{\sth}{s_\text{th}}

\begin{document}
\frenchspacing

\title{\boldmath Khuri-Treiman equations for $\pi\pi$ scattering}

\author{M.~Albaladejo}
\email{miguelalbaladejo@gmail.com}
\affiliation{Departamento de F\'isica, Universidad de Murcia, E-30071 Murcia, Spain}
\author{N.~Sherrill}
\email{nlsherri@indiana.edu}
\affiliation{Center for Exploration of Energy and Matter,
Indiana University, Bloomington, IN 47403, USA}
\affiliation{Physics Department, Indiana University, Bloomington, IN 47405, USA}
\author{C.~Fern\'andez-Ram\'irez}
\affiliation{Instituto de Ciencias Nucleares, Universidad Nacional Aut\'onoma de M\'exico, Ciudad de M\'exico 04510, Mexico}

\author{A.~Jackura}
\affiliation{Center for Exploration of Energy and Matter, Indiana University, Bloomington, IN 47403, USA}
\affiliation{Physics Department, Indiana University, Bloomington, IN 47405, USA}

\author{V.~Mathieu}
\affiliation{Theory Center, Thomas Jefferson National Accelerator Facility, Newport News, VA 23606, USA}

\author{M.~Mikhasenko}

\affiliation{{Universit\"at Bonn,
Helmholtz-Institut f\"ur Strahlen- und Kernphysik, 53115 Bonn, Germany}}

\author{J.~Nys}
\affiliation{Theory Center, Thomas Jefferson National Accelerator Facility, Newport News, VA 23606, USA}
\affiliation{Department of Physics and Astronomy, Ghent University, Belgium}
\affiliation{Center for Exploration of Energy and Matter, Indiana University, Bloomington, IN 47403, USA}
\affiliation{Physics Department, Indiana University, Bloomington, IN 47405, USA}

\author{A.~Pilloni}
\affiliation{Theory Center, Thomas Jefferson National Accelerator Facility, Newport News, VA 23606, USA}

\author{A.~P.~Szczepaniak}
\affiliation{Center for Exploration of Energy and Matter,
Indiana University, Bloomington, IN 47403, USA}
\affiliation{Physics Department, Indiana University, Bloomington, IN 47405, USA}
\affiliation{Theory Center, Thomas Jefferson National Accelerator Facility,
Newport News, VA 23606, USA}

\collaboration{Joint Physics Analysis Center}
\preprint{JLAB-THY-18-2658}

\begin{abstract}
The Khuri-Treiman formalism models the partial-wave expansion of a scattering amplitude as a sum of three individual truncated series,  capturing the low-energy dynamics of the direct and cross channels. We cast this formalism into dispersive equations to study $\pi\pi$ scattering, and compare their expressions and numerical output to the Roy and GKPY equations. We prove that the Khuri-Treiman equations and Roy equations coincide when both are truncated to include  only $S$- and $P$-waves. When higher partial waves are included, we find an excellent agreement between the Khuri-Treiman and the GKPY results. This lends credence to the notion that the Khuri-Treiman formalism is a reliable low-energy tool for studying hadronic reaction amplitudes.
\end{abstract}

\maketitle

\section{Introduction}\label{sec:introduction}
Three-body decays offer a unique window into hadron dynamics. They are an especially useful tool for exploring the hadron spectrum  in the {\it exotic} sectors, where resonances appear that cannot be accurately described by constituent quark models. Some notable examples are the mysterious XYZ peaks observed 
 in three-body decays of heavy quarkonia~\cite{Esposito:2016noz,Lebed:2016hpi,Olsen:2017bmm}. Moreover, new methods have recently been developed that enable direct mapping from three-particle spectra in a finite volume to three-particle scattering amplitudes in the infinite volume, opening the door to lattice QCD calculations~\cite{Hansen:2015zga,Hammer:2017kms,Briceno:2017tce}.
Generally speaking, robust methods for constructing reaction amplitudes that fulfill well-known properties from $S$-matrix theory, such as analyticity, unitarity, and crossing symmetry for the analysis of three-particle final states are mandatory.

One of the main issues posed by the presence of hadrons in any reaction is their final-state interactions, which are formally expressed in terms of the unitarity of the $T$-matrix. In the case of $2 \to 2$ scattering, this effect can often be incorporated by neglecting unitarity in the $t$- and $u$-channels while preserving $s$-channel unitarity. For example, in the case of $\pi\pi$ scattering, the results obtained for the $\sigma$ meson with the Roy equations (or other dispersive approaches that account for the left-hand cut singularities in a non-perturbative way) are very similar to those obtained with approaches that neglect altogether the left-hand cut, or take it into account perturbatively (see for example Refs.~\cite{Caprini:2005zr,Albaladejo:2012te,Pelaez:2015qba} and references therein). Despite crossing symmetry, this is certainly not the case for a $1 \to 3$ decay, where ideally one wants to take into account unitarity in the three possible two-hadron channels in the final state.

In the 1960s, Khuri and Treiman proposed a simple amplitude model to study $K \to 3\pi$ decays~\cite{Khuri:1960zz}. This model is based on the factorization of the scattering amplitude $A(s,t,u)$ into a sum of three functions, each of which depends on a single Mandelstam variable only. Several more studies later appeared that expanded upon this representation of the amplitude~\cite{Bronzan:1963mby,Aitchison:1965zz,Aitchison:1966lpz,Pasquier:1968zz,Pasquier:1969dt,Neveu:1970tn} (see also the recent lectures in Ref.~\cite{Aitchison:2015jxa}). 
For the lowest waves, this approach, which we refer to as the Khuri-Treiman (KT) formalism, can be justified in chiral perturbation theory at lowest order via the so-called reconstruction theorem~\cite{Stern:1993rg,Zdrahal:2008bd,Bijnens:2007pr}. % (see also Ref.~\cite{Bijnens:2007pr}).
 From a broader point of view, as we will discuss in detail below, the KT formalism is a rather simple approach for modeling $A(s,t,u)$. Generally speaking, an infinite sum of $s$-channel partial waves that contain both right-hand cut (RHC) and left-hand cut (LHC) discontinuities is substituted by a finite sum of $s$-, $t$-, and $u$-channel \textit{isobar} amplitudes, each of which exhibit only a RHC structure emanating from unitarity in the respective channel. Although it is known that this model fails to properly account for asymptotic behavior at high energies, it is expected to accurately describe amplitudes in the low-energy regime. For this reason, it has often been used to study meson decays, where the energy range is limited by the decay kinematics.
This formalism has recently been reviewed and applied to various three-body decay channels of light and heavy mesons~\cite{Kambor:1995yc,Anisovich:1996tx,Guo:2015zqa,Guo:2016wsi,Colangelo:2016jmc,Albaladejo:2017hhj,Niecknig:2012sj,Danilkin:2014cra,Isken:2017dkw,Niecknig:2015ija,Pilloni:2016obd,Niecknig:2017ylb}. 

Because the KT formalism has been applied to such a wide range of processes, it is important to establish and validate its range of applicability. This is what we propose to do in this paper. Specifically, we compare the KT model amplitude for $\pi\pi$ scattering with the results of other, arguably more sophisticated dispersive approaches. In the same spirit, we will not strictly enforce elastic unitarity, unlike in Refs.~\cite{Khuri:1960zz,Kambor:1995yc,Anisovich:1996tx,Guo:2015zqa,Guo:2016wsi,Colangelo:2016jmc,Albaladejo:2017hhj,Niecknig:2012sj,Danilkin:2014cra,Isken:2017dkw,Niecknig:2015ija,Pilloni:2016obd,Niecknig:2017ylb}. The manuscript is organized as follows. In Sec.~\ref{sec:amplitudes} we present our notation for the general description of $\pi\pi$ scattering, as well as the form of the scattering amplitude in the KT formalism. In Sec.~\ref{sec:spwaves} we analytically compare the KT and Roy equations~\cite{Roy:1971tc} for the lowest partial waves. We explicitly demonstrate that both formalisms exactly coincide when truncated to $S$- and $P$-waves, complementing  former results discussed in Refs.~\cite{Mahoux:1974ej,Knecht:1995ai,Knecht:1995tr,Ananthanarayan:2000ht}. In Sec.~\ref{sec:numresults} we numerically study the results obtained with KT equations considering also $D$- and $F$-waves, using as an input the parameterization of the $\pi\pi$ scattering shift in Ref.~\cite{GarciaMartin:2011cn}. We compare the results obtained with the latter parameterization and with the results of the GKPY equations discussed in the same work. The conclusions are given in Sec.~\ref{sec:disc}. In the Appendix we specifically discuss the contributions of the LHC to the partial waves.

\section{Amplitudes}\label{sec:amplitudes}
We begin by briefly reviewing the structure of $\pi\pi$ scattering as it relates to our analysis of the KT equations. The general form of the isopin invariant $\pi\pi$ scattering amplitude is determined from the matrix elements of the transition-matrix operator $\hat{T}$:
%\begin{align}
%\label{eq:generalpipi}
% A_{ijkl}(s,t,u) &\equiv \frac{1}{32\pi}\braket{\pi_k(p_3) \pi_l(p_4)}{\hat{T}}{\pi_i(p_1) \pi_j(p_2)} \nonumber \\ 
%& = \delta_{ij}\delta_{kl} A(s,t,u) + \delta_{ik}\delta_{jl} A(t,s,u) + \delta_{il}\delta_{jk}A(u,t,s)~.
%\end{align}
\begin{align}
\label{eq:generalpipi}
\frac{1}{32\pi}\braket{\pi_k(p_3) \pi_l(p_4)}{\hat{T}}{\pi_i(p_1) \pi_j(p_2)} &\equiv A_{ijkl}(s,t,u)  = \delta_{ij}\delta_{kl} A(s,t,u) + \delta_{ik}\delta_{jl} A(t,s,u) + \delta_{il}\delta_{jk} A(u,t,s)~.
\end{align}
The Latin indices in Eq.~\eqref{eq:generalpipi} denote Cartesian isospin indices.\footnote{For a detailed derivation of this decomposition we refer the reader to Refs.~\cite{Shirkov:1969,Lanz:2011kzr}.} The variables $s$, $t$, and $u$ refer to the usual Mandelstam variables. In addition, the invariant amplitude can be expressed in terms of a single function $A(s,t,u)$ and permutations of its arguments by virtue of crossing symmetry. Bose symmetry also requires this function to be symmetric in the second and third variable, $A(s,t,u) = A(s,u,t)$. Our primary interest in the structure $A_{ijkl}(s,t,u)$ is its decomposition into amplitudes of well-defined isospin {in the $s$-channel, $A^{(I)}(s,t,u)$, which is accomplished by means of projectors,
\begin{align}
\label{eq:pipiiso}
A_{ijkl}(s,t,u) & = \sum_{I=0}^{2} P^{(I)}_{ijkl}\, A^{(I)}(s,t,u)~,\\
P^{(0)}_{ijkl} & = \frac{1}{3} \delta_{ij}\delta_{kl}~,\nonumber\\
P^{(1)}_{ijkl} & = \frac{1}{2} \left( \delta_{ik}\delta_{jl} - \delta_{il}\delta_{jk} \right)~,\\
P^{(2)}_{ijkl} & = \frac{1}{2} \left( \delta_{ik}\delta_{jl} + \delta_{il}\delta_{jk} \right) - \frac{1}{3} \delta_{ij}\delta_{kl}~\nonumber.
\end{align}
The amplitudes $A^{(I)}(s,t,u)$ can be written, in turn, as combinations of the amplitude $A(s,t,u)$ and permutations of the arguments by comparing Eq.~\eqref{eq:generalpipi} and Eq.~\eqref{eq:pipiiso},
\begin{equation}\label{eq:AstuI}
\left[\begin{array}{c} A^{(0)}(s,t,u) \\ A^{(1)}(s,t,u) \\ A^{(2)}(s,t,u) \end{array}\right] = 
\left[\begin{array}{ccc} 3 & \phantom{-}1 &  \phantom{-}1 \\ 0 & \phantom{-}1 & -1 \\ 0 & \phantom{-}1 &  \phantom{-}1 \end{array} \right]
\left[\begin{array}{c} A(s,t,u) \\ A(t,s,u) \\ A(u,t,s) \end{array}\right]
\equiv
K 
\left[\begin{array}{c} A(s,t,u) \\ A(t,s,u) \\ A(u,t,s) \end{array}\right]~.
\end{equation}
The left-hand side of Eq.~\eqref{eq:AstuI} can be decomposed into an infinite series of $s$-channel partial-wave amplitudes $t_l^{(I)}(s)$,
\begin{align}
A^{(I)}(s,t,u) & = \sum_{\ell=0}^\infty (2\ell + 1) P_\ell(z_s) t_\ell^{(I)}(s)~, \label{eq:AIpwdecomposition}\\
t_\ell^{(I)}(s) & = \frac{1}{2} \int_{-1}^{+1} {\rm d} z_s P_\ell(z_s) A^{(I)}(s,t(s,z_s),u(s,z_s))~, \label{eq:AIpwdefinition}
\end{align}
where $P_\ell(z_s)$ are Legendre polynomials in the variable $z_s$, the $s$-channel scattering angle, which has the form
\begin{equation}
z_s = z_s(s,t,u) = \frac{t-u}{4p^2(s)}~.
\end{equation}
Additionally, the cross-channel variables under this projection are given by
\begin{align}
t(s,z_s) & = -2p^2(s) (1 - z_s)~,\\
u(s,z_s) & = -2p^2(s) (1 + z_s)~,
\end{align}
where $p^2(s) = (s-4m^2)/4$ is the momentum squared in the center-of-mass frame, and $m$ is the charge-averaged pion mass. The normalization of the partial-wave amplitudes is chosen such that:
\begin{equation}\label{eq:t_eta_delta}
t_\ell^{(I)}(s) = \frac{\eta_\ell^{(I)}(s) e^{2i\delta_\ell^{(I)}(s)} - 1}{2i\sigma(s)}~,
\end{equation}
where $\sigma(s) = \sqrt{1-\frac{4m^2}{s}} = 2p(s)/\sqrt{s}$. The threshold parameters (which are dimensionless with our definitions) for the $S$-waves are defined through:
\begin{equation}\label{eq:t_thres_exp}
\text{Re}\,t_0^{(I)}(s) = a_0^{(I)} + b_0^{(I)} \frac{p^2(s)}{m^2} + \cdots ~.
\end{equation}
}

We now wish to describe the amplitude $A(s,t,u)$ within the KT formalism. This involves truncating the infinite series of $s$-channel partial-wave amplitudes $t_\ell^{(I)}(s)$, which contain both RHC and LHC structure. Truncating this expansion at some maximum orbital angular momentum $\ell_{\text{max}}$ defines an amplitude that is regular in the variables $t$ and $u$. The singularities generated by $t$- and $u$-channel physics can be partially recovered by adding truncated $t$- and $u$-channel series expansions of partial-wave-like functions~\cite{Danilkin:2014cra,Pilloni:2016obd,Guo:2016wsi,Szczepaniak:2015hya},
\begin{align}
\label{eq:KTamp}
A(s,t,u) = \sum_{\ell=0}^{\ell_{\text{max}}}(2\ell+1)P_{\ell}(z_s)p^{2\ell}(s)a^{s}_{\ell}(s) + \left(s\rightarrow t\right) + \left(s\rightarrow u\right),
\end{align}
where the functions $a^{s}_{\ell}(s)$, $a^{t}_{\ell}(t)$, and $a^{u}_{\ell}(u)$ are isobars in the indicated channel. By assumption, the isobars contain only RHC singularities in their respective channel variable. It is important to recognize that the isobars are not independent functions: the symmetry $A(s,t,u)$ = $A(s,u,t)$ implies $a^{t}_{\ell}(t) = (-1)^{\ell}a^{u}_{\ell}(t)$. Note that crossing symmetry is respected in Eq.~\eqref{eq:KTamp} since we take $\ell_{\text{max}}$ to be the same integer for each truncated series. The inclusion of the $p^{2\ell}$ factors enforces the proper behavior of $A(s,t,u)$ near threshold~\cite{Albaladejo:2011bu,Albaladejo:2012sa}. From here, we construct isobars of definite isospin in the $s$-channel and denote them as $\bar{a}_{\ell}^{(I)}(s)$. These can be defined by comparing the KT amplitude in Eq.~\eqref{eq:KTamp} with Eq.~\eqref{eq:AstuI}. 
Specifically, we define:
\begin{equation}
\left[\begin{array}{c}
\bar{a}_\ell^{(0)}(x) \\
\bar{a}_\ell^{(1)}(x) \\
\bar{a}_\ell^{(2)}(x)
\end{array}\right] =
K 
\left[\begin{array}{c}
a_\ell^s(x) \\
a_\ell^t(x) \\
a_\ell^u(x)
\end{array}\right]~,
\end{equation}
where $x = s,t,u$ and the matrix $K$ has been introduced in Eq.~\eqref{eq:AstuI}. We get:
\begin{align}
 A(s,t,u) &=
\sum_{\ell=0}^{\ell_{\text{max}}} (2\ell + 1) P_\ell(z_s) p^{2\ell}(s)\frac{\bar{a}_\ell^{(0)}(s) - \bar{a}_\ell^{(2)}(s)}{3} \nonumber \\
& \quad +
\sum_{\ell=0}^{\ell_{\text{max}}} (2\ell + 1) P_\ell(z_t) p^{2\ell}(t)\frac{\bar{a}_\ell^{(1)}(t) + \bar{a}_\ell^{(2)}(t)}{2} \label{eq:KTbasic}\\
& \quad +
\sum_{\ell=0}^{\ell_{\text{max}}} (2\ell + 1) P_\ell(z_u) p^{2\ell}(u)\frac{\bar{a}_\ell^{(1)}(u) + \bar{a}_\ell^{(2)}(u)}{2} (-1)^\ell \nonumber~,
\end{align}
where $z_t$ and $z_u$ are the $t$- and $u$-channel center-of-mass scattering angles,
\begin{align}
z_t & = z_t(s,t,u) = \frac{s-u}{4p^2(t)}~,\\
z_u & = z_u(s,t,u)=\frac{t-s}{4p^2(u)}~.
\end{align}
For the rest of the paper we will use the condensed notation $z_t(s,z) = z_t(s,t(s,z),u(s,z))$
and the analogous notation for $z_u$. In Eq.~\eqref{eq:KTbasic} it should be understood that $\bar{a}_\ell^{(I)}=0$ unless $\ell + I$ is an even integer so that Bose symmetry is respected. We remind the reader that the specific choice of the nomenclature in Eq.~\eqref{eq:KTbasic} is such that the well-defined isospin amplitudes $A^{(I)}(s,t,u)$ have isobars $\bar{a}_{\ell}^{(I)}(s)$ in the $s$-channel projection. More specifically, inserting Eq.~\eqref{eq:KTbasic} into Eqs.~\eqref{eq:AstuI}, one obtains
\begin{align}
A^{(I)}(s,t,u) &= \sum_{\ell=0}^{\ell_{\text{max}}} (2\ell + 1) P_\ell(z_s) p^{2\ell}(s)\bar{a}_\ell^{(I)}(s) \nonumber\\ 
& \quad + \sum_{\ell=0}^{\ell_{\text{max}}}\sum_{I'}(2\ell+1)P_{\ell}(z_t) p^{2\ell}(t) \frac{1}{2} C_{II'} \bar{a}_{\ell}^{(I')}(t) \nonumber\\
& \quad + \sum_{\ell=0}^{\ell_{\text{max}}}\sum_{I'}(2\ell+1)P_{\ell}(z_u) p^{2\ell}(u) \frac{1}{2} C_{II'} \bar{a}_{\ell}^{(I')}(u)(-1)^{I+I'}~, \label{eq:AIstuKT}
\end{align}
where the coefficients $C_{II'}$ are the matrix elements of
\begin{equation}\def\arraystretch{2.2}
C = \left[\begin{array}{*3{>{\displaystyle}r}}
\frac{2}{3} & \phantom{-}2 & \phantom{-}\frac{10}{3} \\
\frac{2}{3} & \phantom{-}1 & -\frac{5}{3} \\
\frac{2}{3} & -1 & \phantom{-}\frac{1}{3}
\end{array}\right]~.
\end{equation}
We again remind the reader that the functions $\bar{a}_{\ell}^{(I)}(s)$ have only a RHC, and for each of them we write a dispersion relation with an arbitrary number of subtractions $n$,\footnote{Strictly speaking, it is more correct to write $n = n_{\ell,I}$ since the number of subtractions performed can vary for each wave. While this possibility has been explored, the results presented in Sec.~\ref{sec:numresults} consider only the same number of subtractions for all waves.}
\begin{equation}
\bar{a}_\ell^{(I)}(s) = \sum_{j=0}^{n-1}\, \frac{\bar{a}_\ell^{(I)(j)}(s_1)}{j!}(s-s_1)^j
+ \frac{(s-s_1)^{n}}{\pi} \int_{s_\text{th}}^\infty {\rm d}s' \frac{{\rm Im} \bar{a}_\ell^{(I)}(s')}{(s'-s_1)^{n}(s'-s)}~,\label{eq:Abar_DR}
\end{equation}
where $s_\text{th} = 4m^{2}$. For simplicity, all of the subtractions are taken at the same point $s_1 < s_\text{th}$, and the subtraction constants are denoted as $\bar{a}_\ell^{(I)(j)}(s_1)$. Similar dispersion relations are implied for the $t$- and $u$-channel. Inserting this dispersive representation into Eq.~\eqref{eq:AIstuKT}, and the resulting amplitudes $A^{(I)}(s,t,u)$ into the partial-wave projection, Eq.~\eqref{eq:AIpwdefinition}, one obtains the KT representation of the partial-wave amplitudes of well-defined isospin in the $s$-channel,
\begin{align}
(t_{\text{KT}})_\ell^{(I)}(s) &= p^{2\ell}(s) \bar{a}_\ell^{(I)}(s) + \sum_{\ell',I'}\sum_{j=0}^{n-1} (2\ell'+1) C_{II'} \frac{R_{\ell\ell'}^{(j)}(s)}{j!}\bar{a}_{\ell'}^{(I')(j)}(s_1) \nonumber\\
& \quad + \sum_{\ell',I'} (2\ell'+1) C_{II'} \frac{1}{\pi}\int_{s_\text{th}}^{\infty} {\rm d}t' S_{\ell\ell'}^{(n)}(s,t') \frac{{\rm Im} \bar{a}_{\ell'}^{(I')}(t')}{(t'-s_1)^{n}}~. \label{eq:tlI_KT}
\end{align}
The following functions have been introduced above:
\begin{widetext}
\begin{align}
R^{(j)}_{\ell\ell'}(s) & = \frac{1}{2} \int_{-1}^{+1} {\rm d} z\, \left(t(s,z)-s_1\right)^j  P_\ell(z)P_{\ell'}(z_t(s,z))p^{2\ell'}\left(t(s,z)\right) ~,\\
S^{(n)}_{\ell\ell'}(s,t') & = \frac{1}{2} \int_{-1}^{+1} {\rm d} z\, \frac{ \left( t(s,z)-s_1 \right)^n P_\ell(z) P_{\ell'}(z_t(s,z)) p^{2\ell'}\left(t(s,z)\right) }{t'-t(s,z)}~.
\end{align}
\end{widetext}
The functions $R^{(j)}_{\ell\ell'}(s)$ are polynomials in $s$. Furthermore, both $R^{(j)}_{\ell\ell'}(s)$ and $S^{(n)}_{\ell\ell'}(s,t')$ behave as $p^{2\ell}(s)$ for $s \to 4m^2$. In the $s$-channel physical region, Eq.~\eqref{eq:tlI_KT}, together with Eq.~\eqref{eq:Abar_DR}, can be written more compactly as a Roy-like equation,
\begin{equation}\label{eq:KT_final}
(t_{\text{KT}})_\ell^{(I)}(s) = p^{2\ell}(s) \left( P_\ell^{(I)}(s) + \sum_{\ell',I'} \int_{\sth}^{\infty} \text{d}t' \frac{Q_{\ell \ell'}^{I I'}(s,t')}{p^{2\ell'}(t')}\, \im (t_{\text{KT}})_{\ell'}^{(I')}(t')\right)~,
\end{equation}
where we used $\im (t_{\text{KT}})_{\ell}^{(I)}(s) = p^{2\ell}(s) \im \bar{a}_\ell^{(I)}(s)$ for $s>\sth$.
The polynomial term $P_\ell^{(I)}(s)$ and the integral kernels $Q_{\ell\ell'}^{II'}(s,t')$ are given by:
\begin{align}
P_{\ell}^{(I)}(s) &= \sum_{\ell',I'} \sum_{j=0}^{n-1} \bigg( \frac{(s-s_1)^j}{j!} \bar{a}_{\ell'}^{(I')(j)}(s_1) \delta_{\ell \ell'} \delta^{II'} + (2\ell'+1) C_{II'} \frac{R^{(j)}_{\ell\ell'}(s)}{p^{2\ell}(s)}\frac{\bar{a}_{\ell'}^{(I')(j)}(s_1)}{j!} \bigg)~, \label{eq:KT_final_pol}\\
 Q_{\ell \ell'}^{II'}(s,t') &= \frac{1}{\pi (t'-s_1)^{n}}\left[\frac{(s-s_1)^{n}}{t'-s} \delta_{\ell \ell'}\delta^{II'} + C_{II'}(2\ell'+1) \frac{S_{\ell\ell'}^{(n)}(s,t')}{p^{2\ell}(s)} \right]~\label{eq:KT_final_ker}. 
\end{align}
This is our final form of the KT equations for $\pi\pi$ scattering.

\section{\boldmath Analytical comparison of KT and Roy equations for $S$- and $P$-waves}\label{sec:spwaves}

We begin in this section with a brief introduction to the Roy equations before comparing them with the KT equations developed in Sec.~\ref{sec:amplitudes}. The Roy equations have been used extensively to study $\pi\pi$ scattering, with initial studies following Roy's original paper~\cite{Basdevant:1972gk,Basdevant:1973ru,Basdevant:1972uv,Basdevant:1972uu,Pennington:1973xv}, but also more recently in the context of newer data~\cite{Ananthanarayan:2000ht,GarciaMartin:2011cn}. The Roy equations impose, within a given kinematical region, rigorous conditions on the determination of partial-wave amplitudes of definite isospin, based on analyticity and $s-u$ crossing symmetry. In their exact form, the Roy equations couple the infinite set of partial waves. However, any practical application requires a finite truncation, which makes them resemble the KT equations as we will demonstrate. The starting point for deriving these equations is to write down a twice-subtracted fixed-$t$ dispersion relation for these amplitudes. Roy realized these could be written as a matrix equation~\cite{Roy:1971tc}:
\begin{equation}\label{eq:Aroy}
A^{(I)}(s,t,u)  = \sum_{I'}\, ({C}_{st})_{II'}\left[\alpha_{I'}(t) + \beta_{I'}(t)(s-u)\right] + \frac{1}{\pi}\int_{\sth}^\infty \frac{\text{d}x}{x^{2}}\left(\frac{s^{2}}{x^{2}-s^{2}}\delta_{II'}+\frac{u^{2}}{x^{2}-u^{2}}(C_{us})_{II'}\right)\im A^{(I')}(x,t,u(x,t)) ~,
\end{equation}
where we are using the notation in Eq.~\eqref{eq:AstuI}. The column vectors characterizing the $t$-dependent subtraction constants $\alpha_{I'}(t)$ and $\beta_{I'}(t)$ have isospin $0,2$ and isospin $1$ components, respectively. The matrices $(C_{st})_{II'}$ and $(C_{su})_{II'}$ are the crossing matrices and are given explicitly in~Ref.~\cite{Roy:1971tc}.\footnote{They are related to the matrix $C$ used in Sec.~\ref{sec:amplitudes} through $(C_{st})_{II'} = \frac{1}{2}C_{II'}$ and $(C_{us})_{II'} = \frac{1}{2}C_{II'}(-1)^{I+I'}$.} 
The Roy equations themselves are the partial-wave projection of Eq.~\eqref{eq:Aroy}. Using the notation of Ref.~\cite{Ananthanarayan:2000ht}, these can be compactly written as 
\begin{equation}\label{eq:troy}
(t_\text{Roy})_\ell^{(I)}(s) = k_\ell^{(I)}(s) + \sum_{\ell',I'} \int_{\sth}^\infty \text{d}x \, K_{\ell \ell'}^{I I'}(s,x)\, \im\,(t_\text{Roy})_{\ell'}^{(I')}(x)~.
\end{equation}
In this equation, the kinematical functions $k_\ell^{(I)}(s)$ are known linear polynomials in $s$ resulting from projecting over the subtraction polynomials. They contain two parameters; the $S0$- and $S2$-wave scattering lengths $a_{0}^{(0)}$ and $a_{0}^{(2)}$, respectively. The integral kernels $K_{\ell \ell'}^{I I'}(s,x)$ are known as well and are fully documented in Appendix A of Ref.~\cite{Ananthanarayan:2000ht}. Including the results of Refs.~\cite{Martin1966_1,Martin1966_2} with experimental information on the scattering lengths and imaginary parts $\im\,(t_\text{Roy})_{\ell'}^{(I')}(s)$ for $s_{\text{th}}< s < \infty$, the real part of Eq.~\eqref{eq:troy} can be fully determined on the interval $-4m^{2} < s < 60m^{2}$, and analytically continued to the complex plane in a region limited by the Lehmann ellipse~\cite{Ananthanarayan:2000ht}. This implies the Roy equations are particularly useful for studying the resonance properties of low-energy $\pi\pi$ scattering. Indeed, they have proven to be a popular resource in this regard.

Now that we have introduced the Roy equations, we proceed by comparing them with our formalism of the KT equations. If restricted to the elastic region, the Roy equations Eq.~\eqref{eq:troy} can be seen as a closed set of coupled nonlinear integral equations for amplitudes $(t_{\text{Roy}})_\ell^{(I)}(s)$, since each of the former can be written in terms of a single function, the phase shift $\delta_\ell^{(I)}(s)$. The same can be said of the KT equations, but this approach will not be pursued here. We instead consider the Roy and KT equations as integral representations of the analytical partial waves in terms of their discontinuities and subtraction constants. As mentioned in Sec.~\ref{sec:introduction}, the aim in this section is to analytically compare KT and Roy equations when truncating both formalisms to include only $S$- and $P$-waves ($\ell,\ell' = 0,1$). One needs to consider the previous sentence with care: by truncation of the Roy equations to $S$- and $P$-waves, we mean letting the values of $\ell$ and $\ell'$ in Eq.~\eqref{eq:troy} to be either $0$ or $1$. This is \textit{not} the same kind of truncation that is done in Ref.~\cite{Ananthanarayan:2000ht} and related analyses of the Roy equations ~\cite{Basdevant:1972gk,Basdevant:1973ru,Basdevant:1972uv,Basdevant:1972uu,Pennington:1973xv,GarciaMartin:2011cn}. Often what is done in analyzing the Roy equations is to separate the partial waves relevant to a specified low-energy scale, where experimental information is often available, from the partial waves relevant in the high-energy region.  Here, we do not split the integration range in Eq.~\eqref{eq:KT_final} and Eq.~\eqref{eq:troy} to regroup waves based on their energy-dependent relevance. Consequently, what we refer to as the Roy equations in comparison to the KT equations under truncation do not coincide with those of Roy equation analyses such as Ref.~\cite{Ananthanarayan:2000ht} nor with their exact form given in Eq.~\eqref{eq:troy}.

From Eq.\eqref{eq:KTbasic}, the KT amplitude $A(s,t,u)$ with $\ell_{\text{max}} = 1$ has the expansion
\begin{equation}\label{eq:Astu_KTlike}
A(s,t,u) = \frac{\bar{a}^{(0)}_0(s)-\bar{a}^{(2)}_0(s)}{3} 
+ \frac{\bar{a}_0^{(2)}(t)}{2} + \frac{3}{8}(s-u)\bar{a}_1^{(1)}(t)
+ \frac{\bar{a}_0^{(2)}(u)}{2} + \frac{3}{8}(s-t)\bar{a}_1^{(1)}(u)~.
\end{equation}
That the Roy equations imply this KT-like structure for $A(s,t,u)$, with the functions $\bar{a}_{\ell}^{(I)}(s)$ having only a RHC, has been already discussed in Refs.~\cite{Mahoux:1974ej,Knecht:1995ai,Knecht:1995tr,Ananthanarayan:2000ht}. In what follows, we show in full detail that Roy and KT equations under the truncation specified give the same partial-wave amplitudes as in Eq.~\eqref{eq:Astu_KTlike}. To that end, let us consider a single subtraction $n=1$ in each of the KT isobars. In this way, the KT polynomials of Eq.~\eqref{eq:KT_final_pol} are
\begin{alignat}{2}
P_0^{(0)}(s) & = \frac{5}{3} \left( \bar{a}_0^{(0)}(s_1) + 2 \bar{a}_0^{(2)}(s_1) \right) & + \frac{3}{4}(3s - 4m^2) \bar{a}_1^{(1)}(s_1)~,\\  
P_0^{(2)}(s) & = \frac{2}{3} \left( \bar{a}_0^{(0)}(s_1) + 2 \bar{a}_0^{(2)}(s_1) \right) & - \frac{3}{8}(3s - 4m^2) \bar{a}_1^{(1)}(s_1)~,\\  
P_1^{(1)}(s) & = & \frac{3}{2}\bar{a}_1^{(1)}(s_1)~.
\end{alignat}
To lighten the notation, since we are considering only one subtraction in each isobar, we will write in this section $\bar{a}_{\ell}^{(I)(j=0)}(s_1) = \bar{a}_\ell^{(I)}(s_1)$. Although we have made a total of three subtractions, there are only two effective parameters, the subtraction constant $\bar{a}_{1}^{(1)}(s_1)$ and the linear combination $\bar{a}_0^{(0)}(s_1) + 2 \bar{a}_0^{(2)}(s_1)$. This is similar to what happens in Roy equations. Let us now pay attention to the difference between the kernels of the KT and Roy equations,
\begin{equation}
\pi \Delta_{\ell \ell'}^{II'}(s,t') \equiv \frac{p^{2\ell}(s)}{p^{2\ell'}(t')} Q_{\ell \ell'}^{II'}(s,t') - K_{\ell \ell'}^{II'}(s,t').
\end{equation}
Let us recall that these kernels appear under an integral in which the integration variable is $t'$. As can be checked by direct computation, it turns out that each $\Delta_{\ell \ell'}^{II'}(s,t')$ is a linear polynomial in the variable $s$, and hence so is its integral. Thus, we can write
\begin{align}
 (t_{\text{KT}})_{\ell}^{(I)}(s) - (t_{\text{Roy}})_{\ell}^{(I)}(s) &= p^{2\ell}(s) P_{\ell}^{(I)}(s) - k_{\ell}^{(I)}(s) \nonumber\\
& \quad + \sum_{\ell',I'} \int_{\sth}^\infty \text{d}t' \Delta_{\ell \ell'}^{II'}(4m^2,t')\, \im (t_{\text{Roy}})_{\ell'}^{(I')}(t') \nonumber\\
& \quad + (s-4m^2) \sum_{\ell',I'} \int_{\sth}^{\infty} \text{d}t' \left[\frac{\partial}{\partial s}\Delta_{\ell \ell'}^{II'}(4m^2,t')\right]\, \im (t_{\text{Roy}})_{\ell'}^{(I')}(t') \nonumber \\
& \equiv p^{2\ell}(s) P_{\ell}^{(I)}(s) - k_{\ell}^{(I)}(s) + x_\ell^{(I)} + (s-4m^2) y_\ell^{(I)}~,
\end{align}
where the imaginary part $\im (t_\text{KT})^{(I')}_{\ell'}(t')=\im (t_\text{Roy})^{(I')}_{\ell'}(t')$ are given as an input, and 
 $x_\ell^{(I)}$ and $y_{\ell}^{(I)}$ are constants not depending on $s$. Moreover, they are related through:
\begin{equation}
6y_1^{(1)} = -2y_0^{(2)} = y_0^{(0)} = \frac{2x_0^{(0)} - 5 x_0^{(2)}}{12m^2}~.
\end{equation}
The above difference $(t_{\text{KT}})_{\ell}^{(I)}(s) - (t_{\text{Roy}})_{\ell}^{(I)}(s)$ is also a linear polynomial in $s$. The next question is: can we choose our constants $\bar{a}_{1}^{(1)}(s_1)$ and $ \bar{a}_0^{(0)}(s_1) + 2 \bar{a}_0^{(2)}(s_1)$ such that $(t_{\text{KT}})_{\ell}^{(I)}(s) - (t_{\text{Roy}})_{\ell}^{(I)}(s) = 0$ for the three $S$- and $P$-waves? The $P$-wave is proportional to $(s-4m^2)$, so $x_1^{(1)} = 0$. We thus have to satisfy five conditions with only two parameters. It turns out that the solution 
\begin{align}
3 \left( \bar{a}_0^{(0)}(s_1) + 2 \bar{a}_0^{(2)}(s_1) \right) & = \left( a_0^{(0)} + 2 a_0^{(2)} \right) - \left( x_0^{(0)} + 2 x_0^{(2)} \right)~,\\
27 m^2 \bar{a}_1^{(1)}(s_1) & = \left( 2a_0^{(0)} -5a_0^{(2)} \right) - \left( 2x_0^{(0)} -5x_0^{(2)} \right)~,
\end{align}
fulfills indeed the five conditions. Hence, we have proved that, under the conditions indicated above,
\begin{equation}
(t_{\text{KT}})_{\ell}^{(I)}(s) - (t_{\text{Roy}})_{\ell}^{(I)}(s) = 0~.
\end{equation}
This is a highly non-trivial result: the kernels of both approaches are originally different, but it turns out that they are such that the differences in the partial-wave amplitudes are only polynomials, and that the three polynomials can be put to zero with only two free parameters.

For higher waves, $\ell, \ell' \geqslant 2$, the kernels $K_{\ell \ell'}^{II'}(s,t')$ have more complicated structures. For instance, when $\ell' =2$, one also finds the term $\log \left( 1+ \frac{s-4m^2}{2t'} \right)$, in addition to $\log \left( 1+ \frac{s-4m^2 }{t'}\right)$. In that case, the differences in the kernels are not purely polynomials, and the above result cannot be proved. 

\section{Numerical results}\label{sec:numresults}
Since the proof of the result in the previous section only holds when restricting to $S$- and $P$-waves, in this section we numerically study to what extent the KT equations are useful when higher partial waves are considered. Specifically, we now set $\ell_{\text{max}}=3$ in Eq.~\eqref{eq:KTbasic} and thus consider up to the $F$-wave. As stated above, the KT and Roy equations give the real part of the amplitudes once the imaginary part is given, up to a polynomial contribution. As an input for the KT equations, we shall use the CFD parameterization of Ref.~\cite{GarciaMartin:2011cn}, which we now briefly discuss. This work parameterizes the inelasticities $\eta_{\ell}^{(I)}(s)$ and phase shifts $\delta_{\ell}^{(I)}(s)$ of the partial-wave amplitudes, Eq.~\eqref{eq:t_eta_delta}. Two different parameterizations are given in Ref.~\cite{GarciaMartin:2011cn}, called UFD and CFD, which respectively stand for unconstrained and constrained fit to data. They do not differ on the form of the parameterization but in the values the parameters take. In the UFD parameterization only the data are fitted, while in the CFD dispersive constraints are imposed on the amplitudes. Among these, the most relevant ones are the Roy and GKPY equations, which are, respectively, twice- and once-subtracted dispersion relations for the $\pi\pi$ amplitude. Hence, the amplitudes computed with the CFD parameterization satisfy, within uncertainties, these dispersive equations. Both parameterizations of Ref.~\cite{GarciaMartin:2011cn} provide the phase shifts and inelasticities along the RHC. These quantities can be used as inputs to the Roy or GKPY equations so that the amplitudes can be computed at any point on the complex plane. The real part of the amplitudes obtained with the Roy or GKPY equations along the RHC are very similar to those obtained with the CFD parameterization for the amplitudes, since the latter are constrained to satisfy the former. We can use as well the CFD parameterization of Ref.~\cite{GarciaMartin:2011cn} as an input for our KT equations to obtain the real part of the partial-wave amplitudes $(t_{\text{KT}})_\ell^{(I)}(s)$, and compare our results with the original input.

\begin{figure*}
\begin{tabular}{cc}
\includegraphics[height=5.5cm,keepaspectratio]{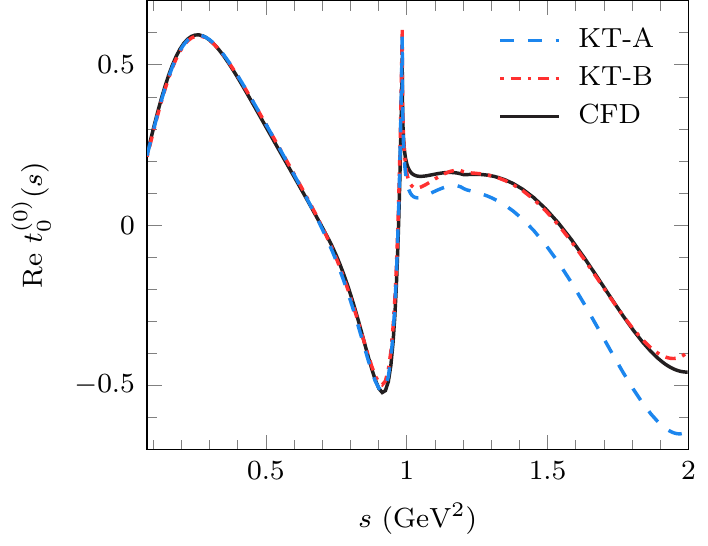} &
\includegraphics[height=5.5cm,keepaspectratio]{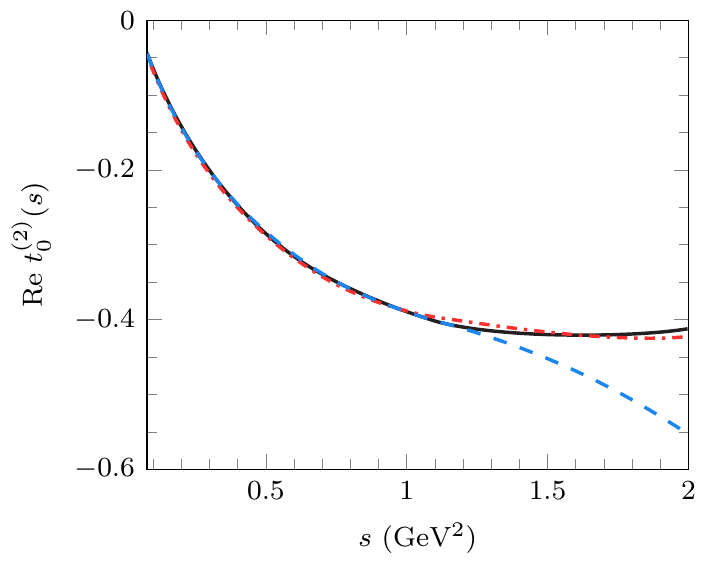} \\
\includegraphics[height=5.5cm,keepaspectratio]{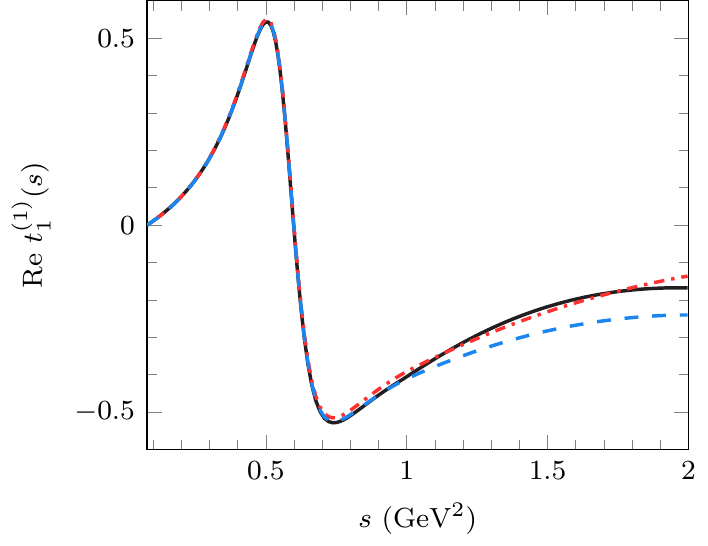} &
\includegraphics[height=5.5cm,keepaspectratio]{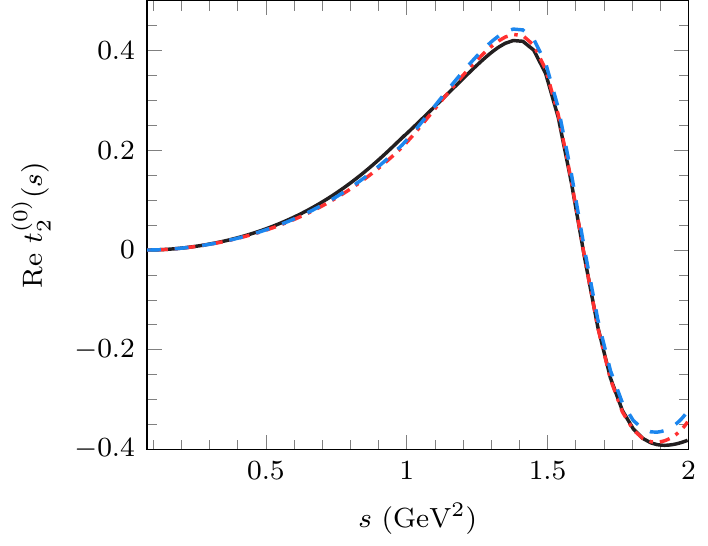} \\
\includegraphics[height=5.5cm,keepaspectratio]{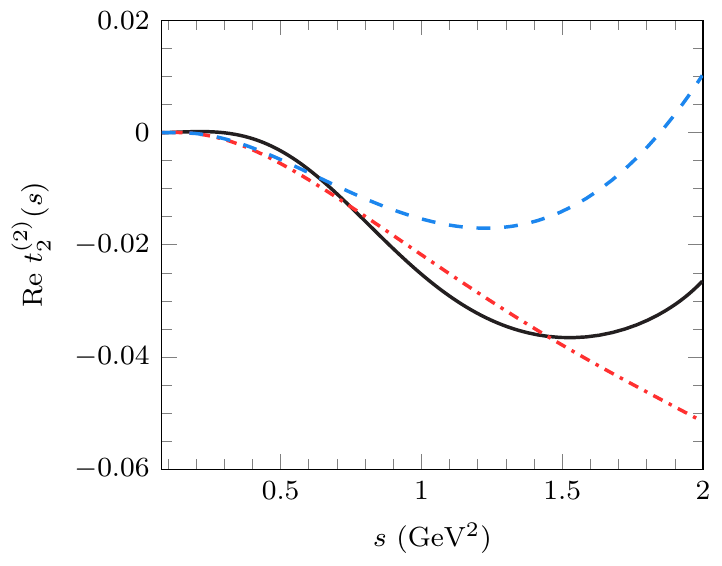} &
\includegraphics[height=5.5cm,keepaspectratio]{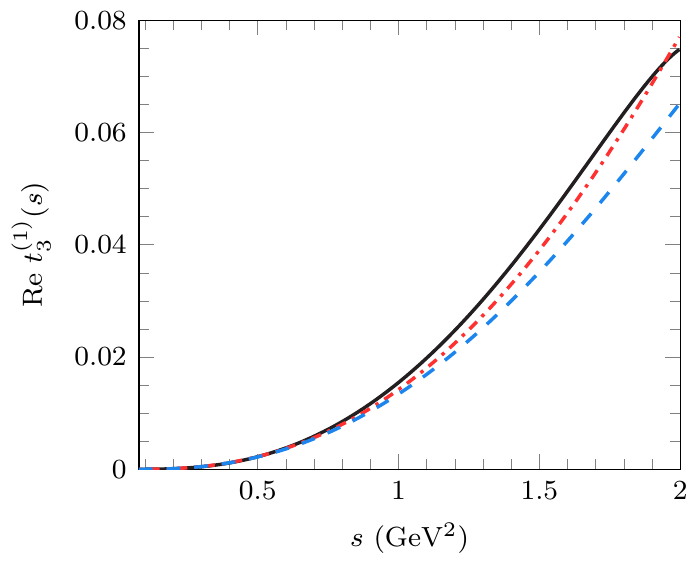}
\end{tabular}
\caption{Comparison of the real part of the different input partial-wave amplitudes $(t_{\text{CFD}})_\ell^{(I)}(s)$ taken from Ref.~\cite{GarciaMartin:2011cn} (black lines, labeled as CFD) with $(t_{\text{KT}})_\ell^{(I)}(s)$. The two setups A (blue dashed line) and B (red dash-dotted line) represent different choices of the number of subtractions in each wave and the maximum value of $s$ ($s_f$) taken into account into the fit of the subtraction constants (see text for more details).\label{fig:comparison}}
\end{figure*}

\begin{figure*}
\includegraphics[height=5.7cm,keepaspectratio]{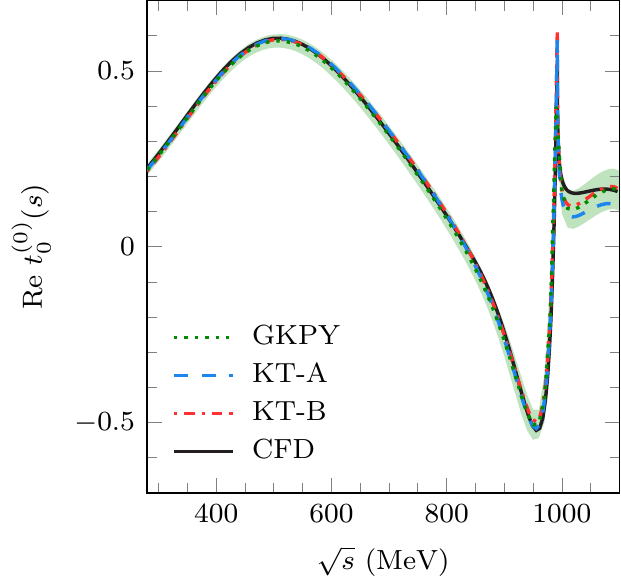}
\includegraphics[height=5.7cm,keepaspectratio]{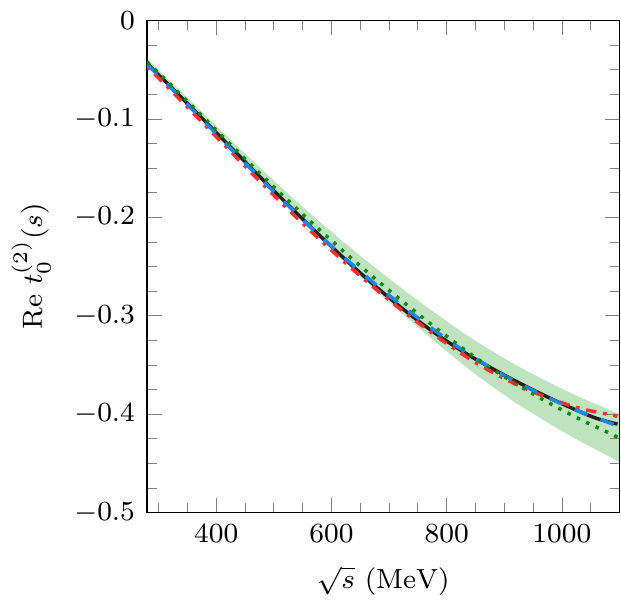}
\includegraphics[height=5.7cm,keepaspectratio]{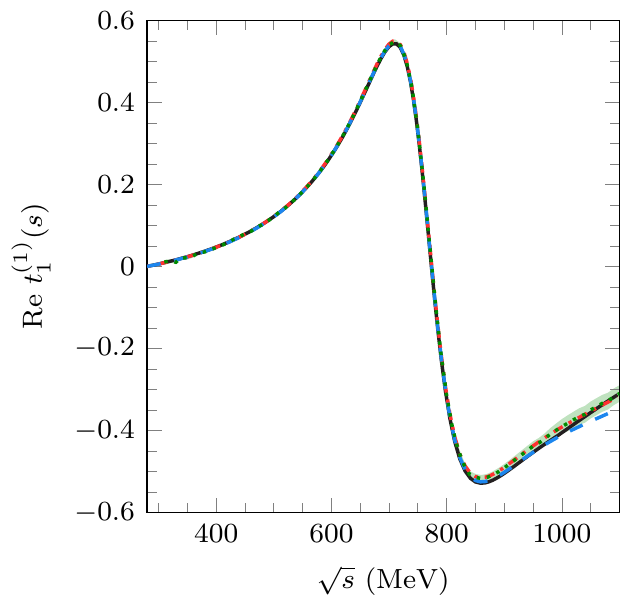}
\caption{Same as Fig.~\ref{fig:comparison}, but restricted to $S$- and $P$-waves, and including the dispersive output (green dotted lines, GKPY) of Ref.~\cite{GarciaMartin:2011cn}. The (green) error band is associated with the GKPY dispersive output, as given in Ref.~\cite{GarciaMartin:2011cn}.\label{fig:comparison2}}
\end{figure*}

The real parts of the input amplitudes are the black solid lines in Fig.~\ref{fig:comparison}. The CFD parameterization reaches up to a center-of-mass energy squared $s_m = (1.42\ \text{GeV})^2 \simeq 2\ \text{GeV}^2$. Since the dispersive integrals in the KT equations extend to infinity, we shall take as an approximation $\delta_\ell^{(I)}(s) = \delta_\ell^{(I)}(s_m)$ and $\eta_\ell^{(I)}(s) = \eta_\ell^{(I)}(s_m)$ for $s \geqslant s_m$. It is important to mention here that we do not use the full CFD parameterization since above $s \simeq 2\ \text{GeV}^2$ we set the phase shift to a constant instead of using the Regge formulas of Ref.~\cite{GarciaMartin:2011cn}. With the input amplitudes fixed, the only remaining freedom in the KT equations are the subtraction constants appearing in the polynomial terms. The subtraction constants are chosen so as to minimize the difference in the region $\sth \leqslant s \leqslant s_f$ between the real part of the input amplitudes and those computed with the KT equations. To be more specific, we are minimizing the following $\chi^2$-like function:
\begin{equation}
\chi^2 = \frac{1}{s_f - \sth}\sum_{\ell,I} \int_{\sth}^{s_f} \text{d}s \left( \re\left[ (t_{\text{KT}})_{\ell}^{(I)}(s) - (t_{\text{CFD}})_{\ell}^{(I)}(s) \right] \right)^2~.
\end{equation}
We recall that the goal here is not to describe the phase shifts and inelasticities parameterized in Ref.~\cite{GarciaMartin:2011cn}, but rather a comparison between the amplitudes used as an input and the output given by the KT equations. For this reason we will not dwell on the calculation of errors, which should be approximately equal to those given by the CFD parameterization. Two different setups for the fits will be considered. In setup A we choose $n_{\ell,I} = 1$ and $s_f = 1\ \text{GeV}^2$, while setup B is computed with $n_{\ell,I} = 2$, and $s_f = 1.9\ \text{GeV}^2 \lesssim s_m$, where $s_m$ has been defined above. In setup A we seek a description of low energies with a small number of subtractions. In contrast, in setup B we have extended the range of the fit up to the maximum allowed by the CFD parameterization and the number of subtractions is increased. The results of the fits are shown in Fig.~\ref{fig:comparison}. We see that the agreement with the original input is quite good and clearly better for setup B. In Fig.~\ref{fig:comparison2}, showing only $S$- and $P$-waves up to $\sqrt{s} = 1.1\ \text{GeV}$, we include in the comparison the dispersive output obtained with the GKPY equations and its associated error band, as given in Ref.~\cite{GarciaMartin:2011cn}. We see that all curves lie well within or at the edges of the error bands.

\begin{table*}
%\begin{center}
\begin{ruledtabular}
\begin{tabular}{lcccl}
             & KT-A & KT-B & \multicolumn{2}{c}{GKPY---CFD}\\ \hline
 $a_0^{(0)}$ & $0.217$ & $0.213$ & $0.221(9)$ &~\cite{GarciaMartin:2011cn}   \\
 $b_0^{(0)}$ & $0.274$ & $0.275$ & $0.278(7)$ &~\cite{GarciaMartin:2011cn}   \\ \hline
 $a_0^{(2)}$ & $-0.044$ & $-0.047$ & $-0.043(8)$ &~\cite{GarciaMartin:2011cn} \\
 $b_0^{(2)}$ & $-0.078$ & $-0.079$ & $-0.080(9)$ &~\cite{GarciaMartin:2011cn} \\ \hline \hline
$\sqrt{s_\sigma}\ (\text{MeV})$      &  $(448,270)$  & $(448,269)$   & ($457^{+14}_{-13}, 279^{+11}_{-7})$ &~\cite{GarciaMartin:2011jx} \\
$\left\lvert g_\sigma \right\rvert \text{GeV}$    & $3.36 $ &   $3.37$ & $3.59^{+0.11}_{-0.13}$ &~\cite{GarciaMartin:2011jx} \\ \hline
$\sqrt{s_\rho}\ (\text{MeV})$    &  $(762.2,72.4)$ & $(763.4,73.5)$ & $(763.7^{+1.7}_{-1.5}, 73.2^{+1.0}_{-1.1})$ &~\cite{GarciaMartin:2011jx} \\
$\left\lvert g_\rho \right\rvert$  & $5.95$  &   $6.01$ & $6.01^{+0.04}_{-0.07}$ &~\cite{GarciaMartin:2011jx} \\ \hline
$\sqrt{s_{f_0}}\ (\text{MeV})$     & $(1000,24)$   &   $(995,26)$ & $(996 \pm 7, 25^{+10}_{-6})$ &~\cite{GarciaMartin:2011jx} \\ 
$\left\lvert g_{f_0} \right\rvert\ (\text{GeV}) $    & $2.4$ &   $2.3$ & $2.3 \pm 0.2$ &~\cite{GarciaMartin:2011jx} \\ \hline
$\sqrt{s_{f_2}}\ (\text{MeV})$   & $(1275.1,89.5)$ &   $(1268.9,86.4)$ & $(1267.3^{+0.8}_{-0.9}, 87 \pm 9)$ &~\cite{Carrasco:2015fva} \\
$\left\lvert g_{f_2} \right\rvert\ (\text{GeV}^{-1})$   & $5.6$ &   $5.5$ & $5.0 \pm 0.3$ &~\cite{Carrasco:2015fva}
\end{tabular}
\end{ruledtabular}
\caption{The $S$-wave scattering lengths and slope parameters, and the pole positions and couplings for the $\sigma$, $\rho$, $f_0(980)$, and $f_2(1270)$ (the latter two are simply denoted above simply as $f_0$ and $f_2$, respectively) obtained with the KT equations (second and third columns for setups A and B, respectively) are shown. For comparison, we also show in the fourth column the same quantities extracted from the GKPY equations or the CFD parameterization of Ref.~\cite{GarciaMartin:2011cn}, as quoted in Refs.~\cite{GarciaMartin:2011cn,GarciaMartin:2011jx,Carrasco:2015fva}.~\label{tab:comparison}}
%\end{center}
\end{table*}
Together with the general comparison of the real part of the input and KT amplitudes, we can more specifically compare the scattering lengths and slope parameters, cf.~Eq.~\eqref{eq:t_thres_exp}, obtained with the two approaches. In Table~\ref{tab:comparison} we show the $a_0^{(I)}$ and $b_0^{(I)}$ parameters computed with the KT equations and compared with those of Ref.~\cite{GarciaMartin:2011cn} (the results labeled as CFD in that work are quoted here). We see that the agreement is also quite good. This is expected since these parameters control the low-energy behavior of the amplitudes, which the KT equations are able to reproduce in the whole energy range ($\sth \leqslant s \leqslant s_m$) considered here.

Up to now we have checked the agreement between the real part of the amplitudes along the RHC obtained from KT equations and those from the CFD parameterization and the GKPY dispersive amplitudes of Ref.~\cite{GarciaMartin:2011cn}. Since the subtraction constants have been fixed so as to minimize the difference between the input amplitudes and the output from the KT equations, this agreement could be seen as natural. One may therefore ask: to what extent does the agreement stand away from the real axis? In this context, the CFD parameterization of Ref.~\cite{GarciaMartin:2011cn} is used in Refs.~\cite{GarciaMartin:2011jx,Carrasco:2015fva} to compute the amplitudes on the complex plane by means of GKPY and Roy equations, as described above. In particular, the position and coupling of poles associated with several resonances are computed in those works. As done with the Roy and GKPY equations, our KT equations allow us to compute the amplitudes at any point on the complex plane. Hence, we now compare the results obtained in Refs.~\cite{GarciaMartin:2011jx,Carrasco:2015fva} with those obtained with the KT equations. To that end, let us first briefly discuss the relation between resonances and amplitudes, mainly to define our notation. Resonances manifest in the amplitudes as poles on the unphysical Riemann sheets which are continuously connected with the physical one on the real axis. To reach the unphysical sheets the amplitudes must be analytically continued. Denoting as $t_{II}(s)$ the amplitude\footnote{Here, for simplicity in the notation, we drop the $\ell$, $I$ scripts notation.} on the second Riemann sheet, we take its customary definition in terms of the amplitude on the physical sheet, $t_I(s)$:
\begin{equation}\label{eq:2rs}
t_{II}^{-1}(s) =  t_{I}^{-1}(s) + 2i \sigma(s)~. 
\end{equation}
Around the pole $s \simeq s_p$,
\begin{equation}
t_{II}(s) \simeq \frac{\tilde{g}_p^2}{s-s_p},
\end{equation}
and we define the coupling $g_p$ of the resonance to the $\pi\pi$ channel in terms of the residue $\tilde{g}_p$ as:\footnote{We choose this particular definition of the coupling to directly compare with the results given in Refs.~\cite{GarciaMartin:2011cn,GarciaMartin:2011jx,Carrasco:2015fva}.}
\begin{equation}
g_p^2 = -16 \pi \frac{2\ell+1}{(4p^2(s_p))^{\ell}} \tilde{g}_p^2~.
\end{equation}
In Table~\ref{tab:comparison} we show the poles and couplings of the different resonances that show up in the amplitudes considered in this work ($S$-, $P$-, $D$-, and $F$-waves). It can be seen that there is an excellent agreement between the determination obtained with the KT equations and those from the dispersive approach of Refs.~\cite{GarciaMartin:2011jx,Carrasco:2015fva}, which use our same input amplitudes but into dispersive equations in principle very different from KT equations.

\begin{figure*}\begin{center}
\includegraphics[height=6.2cm,keepaspectratio]{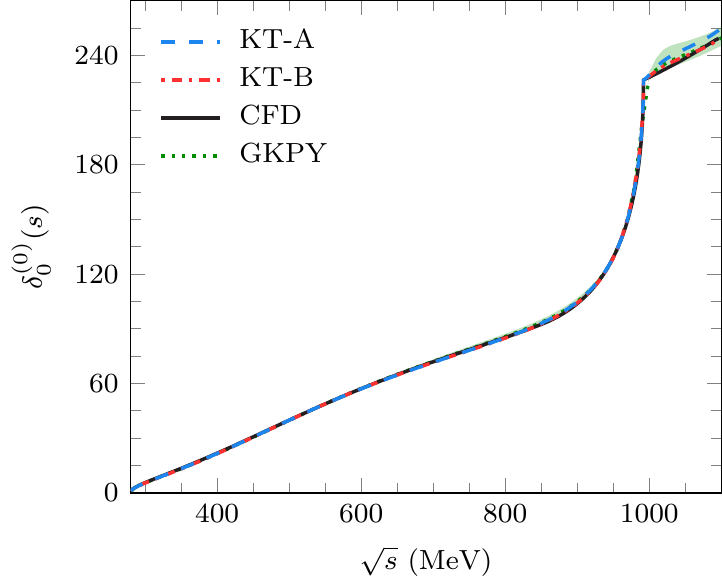} \hspace{1cm}
\includegraphics[height=6.2cm,keepaspectratio]{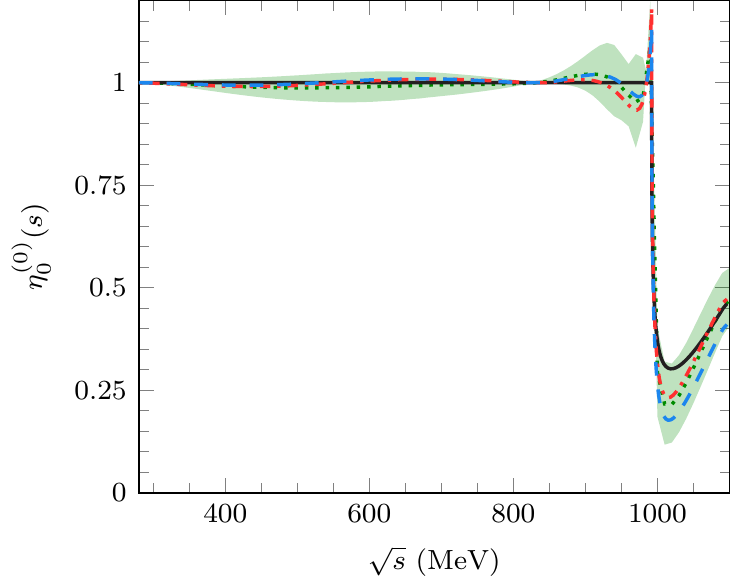}
\caption{Phase shift (left) and inelasticity (right) for the $S0$ wave. The notation for the different curves and the band is as in Fig.~\ref{fig:comparison2}.\label{fig:eta00}}
\end{center}\end{figure*}
Our final discussion about the results for the partial-wave amplitudes obtained with the KT formalism concerns the unitarity of said amplitudes. As explained before, Ref.~\cite{GarciaMartin:2011cn} parameterizes the amplitudes through the phase shifts $\delta_\ell^{(I)}(s)$ and inelasticities $\eta_{\ell}^{(I)}(s)$ as given in Eq.~\eqref{eq:t_eta_delta}, with the latter chosen such that $\eta_{\ell}^{(I)}(s) \leqslant 1$. Hence, the input amplitudes are unitary by construction. However, there is nothing in the KT equations (neither in Roy nor GKPY equations) that force the partial waves to fulfill unitarity---this is only achieved when a specific constraint is imposed for the inelasticity. Hence, unitarity violations\footnote{These refer to either $\eta_{\ell}^{(I)}(s) < 1$ in the elastic region, or $\eta_\ell^{(I)}(s)>1$ in any region.} are to be expected and, indeed, they are present. We take as an example the S0 wave. In Fig.~\ref{fig:eta00} we show the phase shift and inelasticity of this wave as given by the CFD parameterization of Ref.~\cite{GarciaMartin:2011cn} (black solid line), and as obtained from $t_0^{(0)}(s)$ calculated with KT equations (blue dashed and red dash-dotted lines for setup A and B, respectively). Since the real part of $t_0^{(0)}(s)$ closely follows the input amplitude (see Figs.~\ref{fig:comparison} and \ref{fig:comparison2}), and the imaginary parts are equal by construction, we expect $\delta_0^{(0)}(s)$ and $\eta_0^{(0)}(s)$ to be similar in both approaches and, indeed, they are. However, for $\sqrt{s} \leqslant 1\ \text{GeV}$ ($K\bar{K}$ threshold), we see that the input inelasticity is exactly one due to the parameterization. This is not the case for the KT equation, which shows a value $\eta_0^{(0)} \geqslant 1$ in a small region around $\sqrt{s} \simeq 1\ \text{GeV}$. For low energies, $\eta_0^{(0)}$ is very close to 1, and for $\sqrt{s} \geqslant 1\ \text{GeV}$ the KT determination of $\eta_0^{(0)}$ closely follows the inelasticity of the input amplitude. We also show in Fig.~\ref{fig:eta00} $\eta_0^{(0)}$ as computed with the partial wave that results from the GKPY equations (green dotted line), as well as a simple estimation of its error from one of the partial waves (see~Fig.~\ref{fig:comparison2}). We observe that our KT dispersive output lies well within this error band. Furthermore, the value $\eta_0^{(0)}(s)=1$ for $s \leqslant 4m_K^2$ is well comprised in this error band. All this considered, and given that unitarity is not imposed in KT equations, we can safely say that unitarity is well fulfilled for low energies.

For higher energies, $s \to \infty$, due to the $p^{2\ell}(s)$ factors chosen in our definitions of the amplitudes to satisfy the threshold behavior, the absolute value of the real part of the partial-wave amplitudes grows as $s^\lambda$ (with some positive integer $\lambda$), and then the inelasticity does not satisfy unitarity either. This behavior could be corrected by cutting the KT equation for $t_{\ell}^{(I)}(s)$ at some value of $s$ and imposing there some appropriate asymptotic behavior, but this is beyond the scope of our exploratory study. Furthermore, as said, this unwanted behavior occurs for $s \to \infty$, whereas the KT equations are meant to be low-energy approximations. 

\section{Discussion and conclusions}\label{sec:disc}
In this work, we explored the various aspects of the KT formalism within the context of $\pi\pi$ scattering. Our main goal was to assess the validity of the KT formalism within a kinematical range characteristic of hadronic processes in general. Since $\pi\pi$ scattering is the most well studied and simplest, purely hadronic process, this makes it an ideal system for testing the KT formalism. To accomplish this, we derived the KT equations for $\pi\pi$ scattering in Sec.~\ref{sec:amplitudes} and followed this with a proof showing the KT and Roy equations are equivalent when truncating both formalisms to include only $S$- and $P$-waves in Sec.~\ref{sec:spwaves}. While the connection between the Roy equations and the reconstruction theorem under a similar truncation has already been made in previous works, here we demonstrated that a general representation of the amplitude in terms of three distinct expansions in all of the scattering channels also reproduces this result. Numerical results testing the validity of the KT equations including higher waves was explored in Sec.~\ref{sec:numresults}. The dispersive output from the KT equations using as an input the CFD parameterization of Ref.~\cite{GarciaMartin:2011cn} up to the $F$-wave and a center-of-mass energy $\sqrt{s} = 1.42$ GeV was compared with the GKPY equations and the CFD input itself. The KT equations provide an excellent agreement with both the CFD parameterization and GKPY equations at the level of partial-wave amplitudes and subsequent resonance pole parameters. In addition, since KT equations (as well as Roy or GKPY equations) do not imply {\it per se} the unitarity of the partial waves, however we have studied how much they deviate from exact unitarity. We found that the KT equations satisfy unitarity within the CFD parameterization error for low energies. This supports the idea that the KT formalism is a good and simple approach for modeling amplitudes at low energies. The contribution to the partial waves coming from the LHC in the KT approach is also explored in some detail. It is found that, in the scattering region, and for some waves, this contribution can be well accounted for by polynomials.

\appendix*

\section{LHC contributions}\label{app:LHC}
\begin{figure*}
\includegraphics[height=4.5cm,keepaspectratio]{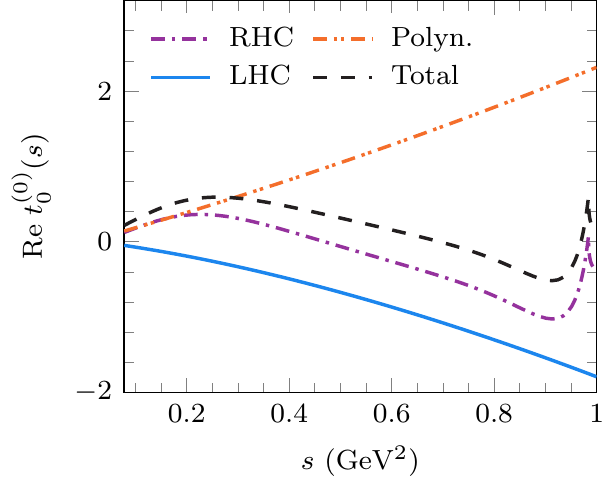}
\includegraphics[height=4.5cm,keepaspectratio]{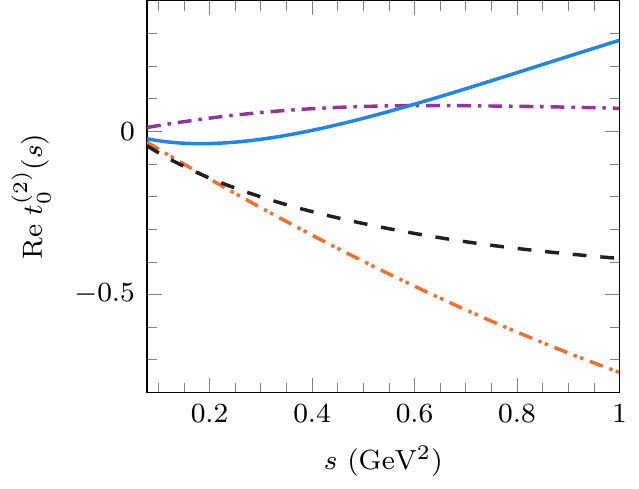}
\includegraphics[height=4.5cm,keepaspectratio]{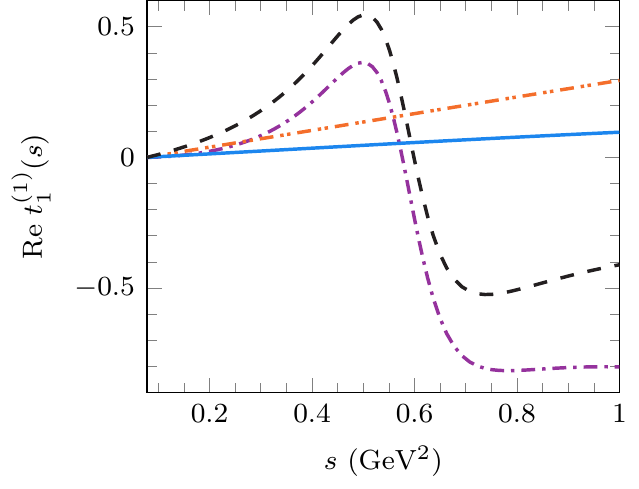}
\caption{Contributions to the total partial-wave amplitude using setup A as described above (black dashed line): RHC integral (purple dash-dotted line), LHC integral (blue solid line), and polynomial term (orange dash-double-dotted line). Only the $S$- and $P$-waves are shown.\label{fig:LHCA}}
\end{figure*}

\begin{figure*}
\includegraphics[height=4.6cm,keepaspectratio]{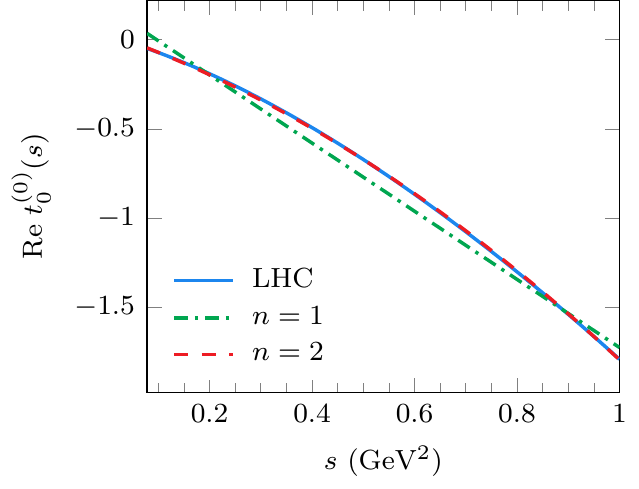}
\includegraphics[height=4.6cm,keepaspectratio]{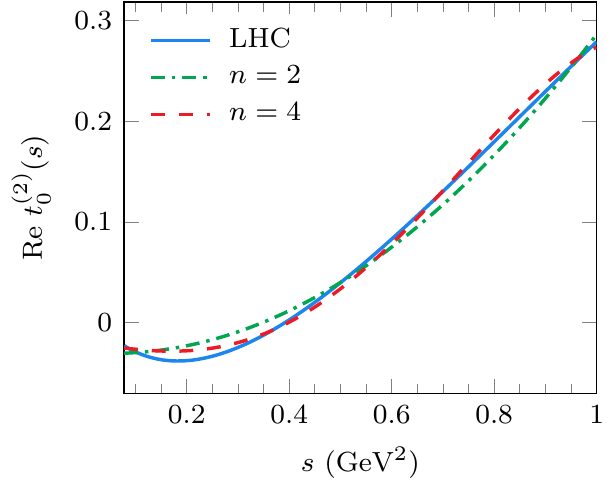}
\includegraphics[height=4.6cm,keepaspectratio]{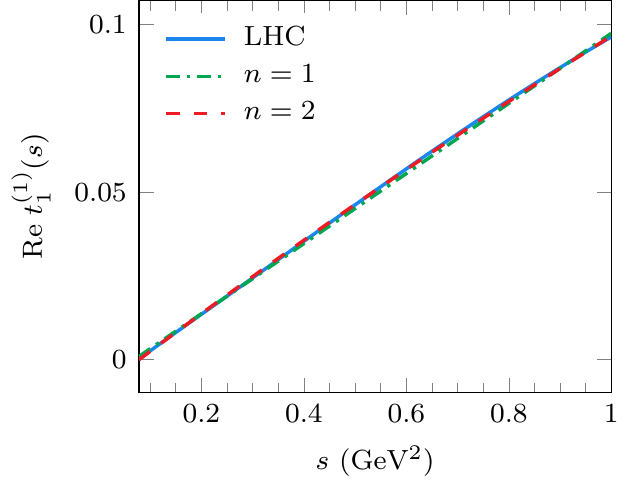}
\caption{Comparison of the LHC integral (blue solid line) shown in Fig.~\ref{fig:LHCA} with order $n$ polynomials (red dashed and green dash-dotted lines) with coefficients fitted to reproduce the LHC integral.\label{fig:LHCB}}
\end{figure*}

A dispersive representation of the $\pi\pi$ partial waves $t_\ell^{(I)}(s)$ would have, generally speaking, three different contributions: two terms from the integrals of the discontinuities along the RHC and LHC, and a polynomial term stemming from the subtractions performed in the dispersion relation. These contributions are easily identifiable in our KT representation of the partial waves, Eqs.~\eqref{eq:KT_final}. The RHC and LHC respectively arise from the singular and non-singular terms of the kernels, Eqs.~\eqref{eq:KT_final_ker}, whereas the polynomial term is given by Eq.~\eqref{eq:KT_final_pol}. In this appendix we discuss the relative importance of the LHC contribution to the partial waves in the KT dispersive representation. In Fig.~\ref{fig:LHCA} we show, for the $S$- and $P$-waves, the three contributions to the partial waves, together with the total amplitude. For simplicity, we consider the number of subtractions given by setup A discussed above for KT approach. We immediately mention two features. First, in the $S0$ and $P$ waves, where prominent resonances show up, the general shape of the amplitude is given by that of the RHC contribution. Second, the LHC contribution is of non-negligible magnitude in the three waves. However, even if the LHC contribution is large, we see that, for the $S0$ and $P$ waves, it has no particular structure in the region $s \leqslant 1\ \text{GeV}^2$ (the maximum range for setup A), whereas it has a more complicated structure in the $S2$ wave. In Fig.~\ref{fig:LHCB} we show the LHC contribution for the three waves compared with order $n$ polynomials, whose coefficients have been fitted to reproduce the LHC contribution. For the $S2$ wave, although the polynomial is able to account for the bulk of the LHC contribution, the particular details of the former cannot be described even with higher-order polynomials. Although this agrees with the expectation for the LHC to dominate the $S2$ lineshape, in absence of any resonance pole, the deviations of the polynomial from the LHC are of the same order as the deviations of KT from the Roy result shown in Fig.~\ref{fig:comparison}, and we cannot derive any strong conclusions from that. On the other hand, for the $S0$ and $P$ waves it can be seen that it is possible to accurately reproduce the LHC contribution with low-order polynomials. Therefore, if one writes down independent (\ie, crossing symmetry violating) dispersion relations for the partial waves $t_{\ell}^{(I)}(s)$, the contributions from the LHC can be reabsorbed into the polynomial coefficients that are already present in the dispersion relation for the RHC. This is the meaning of the common statement that LHC contributions can be neglected (or at least treated perturbatively) in scattering process like $\pi\pi \to \pi\pi$.

\begin{acknowledgments}
MA and NS contributed equally to this work.
MA thanks J.~R.~Pel\'aez, A.~Rodas, and J.~Ruiz de Elvira for useful discussions and for providing numerical results concerning Ref.~\cite{GarciaMartin:2011cn}. NS thanks E.~Passemar for useful discussions.
This work was supported by BMBF,
by the U.S.~Department of Energy under grants No.~DE-AC05-06OR23177 and No.~DE-FG02-87ER40365, by
PAPIIT-DGAPA (UNAM, Mexico) grant No.~IA101717,
CONACYT (Mexico) grant No.~251817, by
Research Foundation -- Flanders (FWO),
by U.S.~National Science Foundation under award Nos.~PHY-1415459 and PHY-1205019,
and Ministerio de Econom\'ia y Competitividad through grants 
Nos.~FPA2016-77313-P, FIS2014-51948-C2-1-P and SEV-2014-0398.
\end{acknowledgments}

\bibliographystyle{apsrev4-1}
\bibliography{quattro}

\end{document}